\documentclass[twoside,onecolumn]{revtex4}
\usepackage{bm}
\usepackage[utf8]{inputenc}
\usepackage{amssymb,theorem,xspace}
\usepackage{float}
\usepackage{amsmath}
\usepackage{mathrsfs}
\usepackage{pgfplots}
\usepackage{graphicx}
\usepackage{subfigure}
\usepackage{pgfplots}
\pgfplotsset{compat=1.17}

\begin{document}
\title{Quark-Antiquark Effective Potential in Symplectic Quantum Mechanics}
\author{R. R. Luz}
\email{renatorodriguesgm@gmail.com}
\affiliation{ International Center of Physics, Instituto de Física, Universidade de Brasília, 70.910-900, Brasília, DF, Brazil.}
\author{Caroline S. R. Costa}
\email{carolsilvarc@gmail.com}
\affiliation{Instituto de F\'{i}sica Te\'{o}rica, Universidade Estadual Paulista, Rua Dr.  Bento Teobaldo Ferraz, 271 - Bloco II, 01140-070,
S\~{a}o Paulo, SP, Brazil.}
\author{G. X. A. Petronilo}
\email{gustavo.petronilo@aluno.unb.br}
\affiliation{ International Center of Physics, Instituto de Física, Universidade de Brasília, 70.910-900, Brasília, DF, Brazil.}
\author{A. E. Santana}
\email{a.berti.santana@gmail.com}
\affiliation{ International Center of Physics, Instituto de Física, Universidade de Brasília, 70.910-900, Brasília, DF, Brazil.}
\author{R. G. G. Amorim}
\email{ronniamorim@gmail.com}
\affiliation{ International Center of Physics, Universidade de Brasília, Faculdade Gama, 72.444-240, Brasília, DF, Brazil.\\Canadian Quantum Research Center, 204-3002 32 Ave Vernon, BC, Canada.}
\author{R. A. S. Paiva}
\email{rendisley@gmail.com}
\affiliation{International Center of Physics, Universidade de Brasília, Faculdade Gama, 72.444-240, Brasília, DF, Brazil.}
\begin{abstract}
In this paper, we study within the structure of Symplectic Quantum Mechanics a bi-dimensional non-relativistic strong interaction system which represent the bound state of heavy quark-antiquark, where we consider a Cornell potential which consists of Coulomb-type plus linear potentials. First, we solve the Schrödinger equation in the phase space with the linear potential. The solution (ground state) is obtained and analyzed by means of the Wigner function related to Airy function for the $c\overline{c}$ meson. In the second case, to treat the Schrödinger-like equation in the phase space, a procedure based on the Bohlin transformation is presented and applied to the Cornell potential. In this case, the system is separated into two parts, one analogous to the oscillator and the other we treat using perturbation method. Then, we quantized the Hamiltonian with the aid of stars operators in the phase space representation so that we can determine through the algebraic method the eigenfunctions of the undisturbed Hamiltonian (oscillator solution), and the other part of the Hamiltonian was the perturbation method. The eigenfunctions found (undisturbed plus disturbed) are associated with the Wigner function via Weyl product using the representation theory of Galilei group in the phase space. The Wigner function is analyzed, the non-classicality of ground state and first excited state is studied by the non-classicality indicator or negativity parameter of the Wigner function for this system. In some aspects, we observe that the Wigner function offers an easier way to visualize the non-classic nature of meson system than the wavefunction does.
\end{abstract}
\maketitle

\section{Introduction}
\label{sec:Introduction}
With the discovery of the $J/\Psi$ meson in 1974 \cite{A_71}, the search to describe such heavy quark systems was based on the approach of potential models. The $J/\Psi$, which has a mass around 3.5 times that of the proton, is the lowest bound state of a charm and anti-charm quark \cite{A_72}. The $c\overline{c}$ meson appears to be an excellent device for testing QCD predictions.
In quantum chromodynamics (QCD), the computation of meson properties is completely non-perturbative. As a result direct calculation utilizing lattice QCD (LQCD) techniques is the only general method accessible. However, other strategies are effective for heavy quarkonia. The light quarks in a meson, the mass of the bound state is much larger than the mass of the quark, move at relativistic speeds. On the other hand, the speed of the charm and bottom quarks in their respective quarkonia is tiny enough that relativistic effects in these states are greatly diminished. The velocity, $\mathbf{v}$, is believed to be around $0.3c$ for charmonia. An extension in powers of $\dfrac{v^2}{c^2}$ can then be used to approximate the computation. This method is known as non-relativistic QCD (NRQCD).

The masses of heavy quarkonium states can be calculated using effective potential models. The non-relativistic motion of the quarks that make up the quarkonium state is exploited in this technique to imply that they move in a static potential, comparable to non-relativistic hydrogen atom models. One of the most prominent potential models is the Cornell (or funnel) potential~\cite{A_0}. Specifically the system is characterized by a linear combination of the Coulomb and linear potentials. The study of these potentials should take into account the two important features of the Quantum Chromodynamics (QCD),
namely, asymptotic freedom and quark confinement~\cite{A_1,A_2,A_3,A_4}. Another interest is that the Cornell potential can be used to analyze the transition between the confined and unconfined phases of matter~\cite{A_5,A_6,A_7}.   
This potential is of considerable importance in various branches of physics such as propagation of gravitational waves, particle and nuclear physics~\cite{A_13,A_56,A_57,A_58,A_60}, mathematical modeling of the parton vibrations inside hadronic system~\cite{A_59,A_52,A_53,A_54,A_55}, quantum chromodynamics and atomic physics~\cite{A_2,A_6,A_13,A_60}. It takes the form:
\begin{equation}\label{cornell1}
    V(q)=\frac{b}{q}+aq,
\end{equation}
where the first term is responsible for the interaction between quark and its antiquark by gluon exchange at short distances. The second term is responsible for quark confinement at large distances. It is known that the potential $V(q)$ given in Eq. \eqref{cornell1} reproduces two specific features of the strong interaction. The Coulomb-like term displays a property known as asymptotic freedom. This phenomenon tells us that strong interaction coupling constant is a function of momentum transfer. At short distances, the momentum transfer increases in a quark-antiquark collision. In this regard, the coupling  constant becomes so small that quarks and gluons can be considered as approximately free, and their interactions can be treated by a perturbation theory~\cite{A_61,A_62,A_63,A_64,A_65,A_68}. Over extended distances the momentum 
transfer decreases while the strong interaction coupling constant becomes larger. On the other extreme, it features a mechanism called confinement that keeps quarks and antiquarks permanently within hadrons at 1 Fermi separations~\cite{A_62,A_65,A_66,A_67}. Confinement is explained consistently in the flux tube model. The gluon field between a pair of color charges develops a string-like configuration as the distance between quarks and antiquarks is extended (flux tube). As a result, across extended distances, the energy density in the tube between the quarks maintaining the gluon field remains constant~\cite{A_62,A_65,A_69}. The energy stored in the field is then proportional to the quark separation, giving us the linear term in the potential of Eq.~\eqref{cornell1}. In that way, at relatively larger separation the potential energy can create new quarks pairs in colorless forms instead of a free quark. The confinement has been a mystery until now since no one has been able to confirm it. Although there has been recent progress using the techniques of lattice QCD~\cite{A_62,A_65,A_67}.

There are several studies with the Cornell potential in literature. For instance, Vega e Flores~\cite{A_6,A_8} studied the Schrödinger equation with the Cornell potential using variational method and supersymmetric quantum mechanics~\cite{A_6,A_7,A_8,A_9,A_10}. Bruni $\textit{et al}$~\cite{A_5} calculated the energy configuration for a quark-antiquark pair from the Nambu-Goto action. It was shown that this configuration energy has the shape of a Cornell potential~\cite{A_5}. Khoka $\textit{et al}$~\cite{A_9,A_11} utilized the analytical exact iterative method to solve N-dimensional Schrödinger equation with an extended Cornell potential~\cite{A_9}. They found the energy and mass spectrum of heavy quarkonia~\cite{A_9}. Omugbe~\cite{A_9} applied the Nikiforov-Uvarov method to obtain the eigensolutions of the radial Schrödinger equation with Cornell potential plus an inversely quadratic potential~\cite{A_9,A_12,A_13}. Until now, an analysis of such states in quantum phase space via the Wigner function is still lacking. In this work, we investigate  confinement term of strong interaction system described by heavy quarks-antiquarks in the phase space.

The first successful formalism of quantum mechanics in phase space was introduced by Wigner in 1932~\cite{A_14,A_15,A_16,A_17,A_18,A_19,A_20,A_21,A_22,A_23,A_24,A_25,A_26,A_27,A_28,A_29,A_30,A_31,A_32,A_33,A_34,A_35,A_36}. He was motivated by the problem of finding a way to improve the quantum statistical mechanics~\cite{A_21,A_22,A_23,A_24,A_25,A_26,A_29,A_30,A_31,A_32,A_33,A_37,A_38,A_39,A_40,A_41,A_42,A_43,A_44,A_45,A_46,A_47,A_48,A_49}. In the Wigner formalism an operator $A$ defined in Hilbert space $\mathcal{H}$, is associated to a function in a phase space, say $A\rightarrow a_{W}(q,p)$. The product of two operators, $AB$, both defined in $\mathcal{H}$, is associated to a Weyl product or star-product of their phase space correspondents, i.e.~\cite{A_40,A_41,A_42,A_43,A_44,A_45,A_46,A_47,A_48},
\begin{equation}
    a_{W}\star b_{W}=a_{W}(q,p)e^{\frac{i \Lambda }{2}}b_{W}(q,p), \label{L_1}
\end{equation}
where $\Lambda =\overleftarrow{\partial}_{p}\overrightarrow{\partial}_{q}-\overleftarrow{\partial}_{q}\overrightarrow{\partial}_{p}$. Such result can be seen as the action of a star-operator, $a_{W}(q,p)\star$, in the function $b_{W}(q,p)$. This fact is important for define the star operators $a_{W}(q,p)\star$, which are used to build up representations of symmetry groups in a symplectic manifold~\cite{A_29,A_30,A_31,A_43}. This gives rise, for example, to the Klein-Gordon and Dirac equations written in phase space~\cite{A_23,A_40,A_41,A_43}. The connection with Wigner function is derived, providing a physical interpretation for the representation. These symplectic representations provide a way to consider a perturbative approach for Wigner function on the bases of symmetry groups.

The aim of this work is to analyze the behavior of the Wigner function given in terms of Airy function for the ground state of $c\overline{c}$ meson in the formalism of the symplectic quantum mechanics. For this, we considering only the linear potential that is a model more simple of the Cornell potential. In sequence, we studied the Schrödinger equation in the phase space to obtain the energy eigenvalues and corresponding wave functions by using the Bohlin transformation and perturbation theory for the Cornell potential. Then, we calculate the Wigner function for the heavy $c\overline{c}$ meson ($J\psi$). Consequently, we calculate the negativity parameter of the Wigner function associated to the ground state and first excited state of $c\overline{c}$ meson that present nonclassical behaviour. The study of the Wigner function for such a system is crucial in order to learn more about the system's chaotic nature.

The paper is organized as follows. In section~\ref{sqm}, we write the Schrödinger equation in phase space and we perform the relation between phase space amplitude and Wigner function. In section~\ref{linear}, we solve the Schrödinger equation in phase space for the quark-antiquark potential and calculate the Wigner function. In section~\ref{galilean}, the perturbative method is presented in order to solve the Schrödinger equation in phase space with the Cornell potential, the Wigner function is calculated. The conclusion is presented in section~\ref{sec:Conclusion}.

\section{Symplectic Quantum Mechanics: outline and notation}\label{sqm}

In this section we construct a formalism to quantum mechanics in phase space. For this purpose, we introduce an Hilbert space associated to phase space, denoted by $\mathcal{H}(\Gamma)$. In this sense,  the association of the Hilbert space $\mathcal{H}$ with the phase space $\Gamma$ is given by
$\int dpdq\; \phi^{\ast}(q,p)\phi(q,p) <\infty$ where $\phi(p,q)\in\Gamma$. Then we write $\phi(q,p)=\langle q,p|\phi\rangle$, with the aid
$\int dp dq |q,p\rangle\langle q,p| =1$
such that $\langle\phi|$ is the dual vector of $|\phi\rangle$. This symplectic Hilbert space is denoted by $H(\Gamma)$.
To construct a symplectic representation of quantum mechanics we define the momentum and position operator
\begin{eqnarray}
    \widehat{P}=p\star=p-i\frac{\partial}{\partial q}, \label{g1}\\
    \widehat{Q}=p\star=q+i\frac{\partial}{\partial p}. \label{g2}
\end{eqnarray}
Which satisfy the Heisenberg commutation relation $\left [ \widehat{Q},\widehat{P} \right ]=i$. 

Introduced the following operators
\begin{eqnarray}
    \widehat{K}_i&=&m\widehat{Q}_i-t\widehat{P}_i,\\
    \widehat{L}_i&=&\epsilon_{ijk}\widehat{Q}_j\widehat{P}_k,\\
    \widehat{H}&=&\frac{\widehat{P}^2}{2m},
\end{eqnarray}
we obtain the following set of commutation rules:
\begin{equation}
    \begin{aligned}
    &\left[\widehat{L}_i,\widehat{L}_j\right]=i\epsilon_{ijk}\widehat{L}_k,\qquad\left[\widehat{L}_i,\widehat{k}_j\right]=i\epsilon_{ijk}\widehat{K}_k, \qquad \left[\widehat{K}_i,\widehat{H}\right]=i\widehat{P}_i,\\
    &\left[\widehat{L}_i,\widehat{P}_j\right]=i\epsilon_{ijk}\widehat{P}_k,\qquad\left[\widehat{K}_i,\widehat{P}_j\right]=i m\delta_{ij}\mathbf{1},\\ 
    \end{aligned}
\end{equation}
with all the other commutations being zero. This is the Galilei-Lie algebra with a central extension characterized by $m$, where $\widehat{P},\widehat{K},\widehat{L}$ and $\widehat{H}$, are the generators of translation, Galilean boosts, rotation and time translation respectively.

The evolution of the wave function in phase space is
$$
\psi(q,p,t)=e^{\widehat{H}t}\psi(q,p,0),
$$
derivating with respect to time we obtain
\begin{eqnarray*}
\partial_t \psi(q,p;t)&=&\left (\frac{\widehat{P}^2}{2m}+V(\widehat{Q})  \right )\psi(q,p;t),\\
\end{eqnarray*}
this is the Schrodinger-like equation in phase space~\cite{A_33}.

The association of $\psi(q,p,t)$ with a function $f_w$ is giving by~\cite{A_33}
$$
f_w(q,p,t)=\psi(q,p,t)\star\psi^\dagger(q,p,t),
$$
which has all the properties of the Wigner function~\cite{A_70}. In the next section, we solve the representation Schrödinger equation in phase space and establish the relation of amplitudes in phase space and the Wigner function for the $c\overline{c}$ meson system.

\section{Quark Confinement and Schrödinger equation in Phase Space }\label{linear}
In this section, we analyze the quark confinement in the context of phase-space. For this purpose, we consider the second term of Cornell potential given in Eq.(\ref{cornell1}). Therefore, we find a solution for the Scrödinger equation in phase space with linear potential and construct the associated Wigner function. We write the Schrödinger equation in phase space with the linear potential $V(q)$ as follows
\begin{equation}
    \frac{p^2}{2m}\star \psi(q,p) + \lambda q\star \psi(q,p) = E\psi(q,p). \label{D_0}
\end{equation}

\begin{figure}[H]
    \centering
   \includegraphics[scale=0.38]{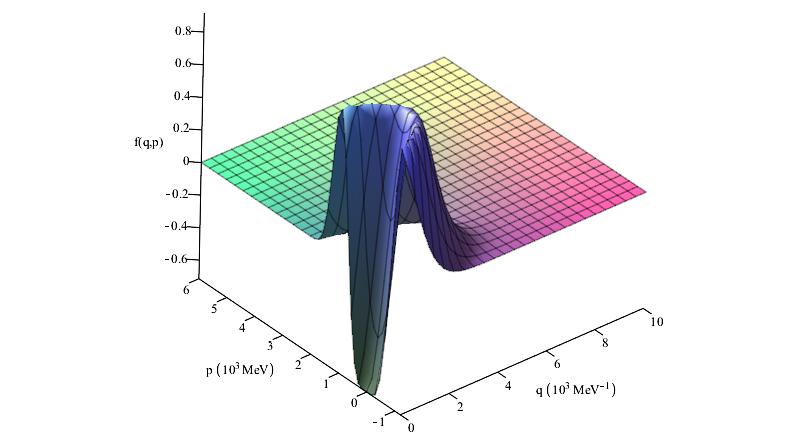}
    \caption{Wigner function for ground state of the $c\overline{c}$ meson.}
    \label{fig:wigner}
\end{figure}

\begin{figure}[!ht]
\centering
\includegraphics[scale=0.38]{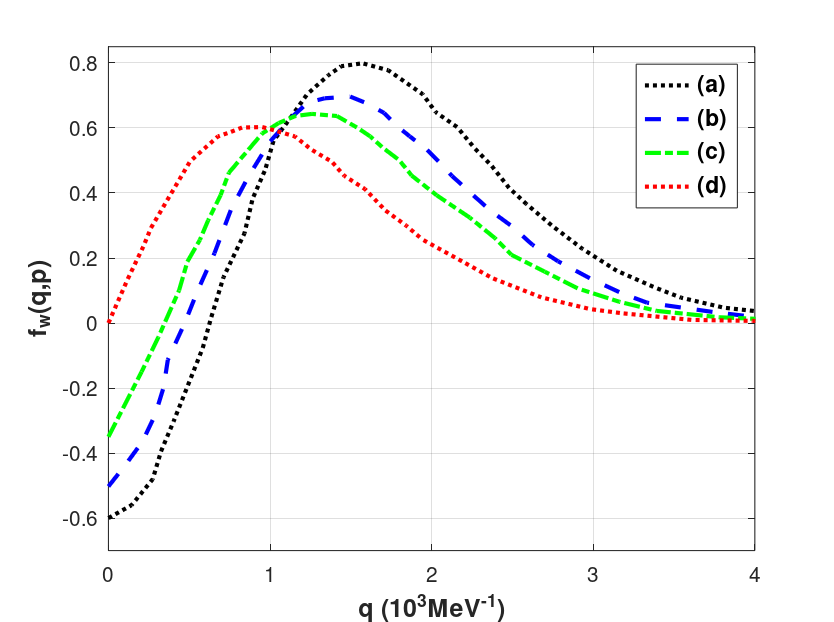}
\caption{The curves $(a)$-$(d)$ represents cut off graphics of the Fig. \ref{fig:wigner} for the ground state $c\overline{c}$ meson with varying of the kinetic energy. }
  \label{fig: wigner1}
\end{figure}

And using the equations (\ref{g1}), (\ref{g2}) in (\ref{D_0})  we get the Schrödinger equation in the form
\begin{equation}
    \frac{1}{2m}\left ( p^{2}-i p\partial_{q}-\frac{1}{4}\partial^{2}_{q} \right )\psi + \lambda\left ( q+\frac{i}{2}\partial_{p} \right )\psi=E\psi, \label{D_3}
\end{equation}
where we used $\hbar=1$. By using the transformation $\omega =\frac{p^2}{2m} + \lambda q$, such that
\begin{equation}
    \frac{\partial \psi}{\partial q}=\frac{\partial \psi}{\partial \omega}\frac{\partial \omega}{\partial q}=\lambda\frac{\partial \psi}{\partial \omega}, \label{D_4}
\end{equation}
with this we obtain
\begin{equation}
    \alpha \frac{\partial^2 \psi}{\partial \omega^2} - \omega\psi=-E\psi, \label{D_6}
\end{equation}
where $\alpha =\frac{\lambda^2}{8m}$. Then, the Schrödinger equation is reduced to the Airy equation~\cite{A_51}
\begin{equation}
    \frac{\partial^2 \psi}{\partial \omega^2}-\kappa \psi=0. \label{S_2}
\end{equation}
Where $\kappa = \frac{\omega-E}{\alpha }$. The solution of this equation is
\begin{equation}
    \psi(\omega)=C_{1}Ai\left [ \alpha ^{2/3}\left ( \frac{\omega}{\alpha}-\frac{E}{\alpha} \right ) \right ]+C_{2}Bi\left [ \alpha^{2/3}\left ( \frac{\omega}{\alpha}-\frac{E}{\alpha} \right )  \right ].
\end{equation}
Where $Ai$ and $Bi$ are homogeneous Airy functions. However $Bi$ goes to infinity for $\omega \rightarrow \infty $, this solution is not relevant, so $C_{2}=0$~\cite{A_51}. In this case, the solution of the Eq. \eqref{S_2} is reduced to~\cite{A_51}
\begin{equation}
    \psi(q,p)= C_{1}A_{i}\left [ \left ( \frac{\lambda^2}{8m} \right ) ^{-1/3}\left ( \frac{p^2}{2m}+\lambda q-E \right ) \right ],
\end{equation}
where $\psi(q,p)$ is a real function. As consequence, by virtue of associativity, $\psi \star \psi \propto \psi$, we write the solution of Wigner function in terms of the Airy function
\begin{equation}
    f_{w}(q,p)=NA_{i}\left [ \left ( \frac{\lambda^2}{8m} \right )^{-1/3}\left ( \frac{p^2}{2m}+\lambda q-E \right ) \right ].
\end{equation}
Where $N$ is a normalization constant. By condition $f_{w}(q,p)=0$, we can determine the energy levels
\begin{equation}
    Ai\left [ \left ( \frac{ \lambda^2}{8m} \right )^{-1/3}\left ( \frac{p^2}{2m}+\lambda q-E \right ) \right ]=0.
\end{equation}
So, the energy of system are determined by 
\begin{equation}
  E=\frac{p^2}{2m}-\alpha^{1/3}r_{i},  \label{G_0}
\end{equation}
where $r_{i}$ is the zero of the Airy function. In our case, we consider only the energy for ground state
\begin{equation}
    E=\frac{p^2}{2m}-\left ( \frac{\lambda^2}{8m} \right )^{1/3}r_{0}.
\end{equation}
In consequence, the normalization factor see Ref. \cite{A_51}, is determined by 
\begin{equation}
 \int_{0}^{\infty}\int_{-p}^{p}f_{w}(q,p)dqdp=1,    
\end{equation}
then we get
\begin{equation}
    f_{w}(q,p)=NA_{i}\left [ \left ( \frac{\lambda^2}{8m} \right )^{-1/3}\left ( \frac{p^2}{2m}+\lambda q-E \right ) \right ]. \label{G_3}
\end{equation}
Where $N=\alpha^{-1/3}$ and $m$ is the reduced mass of the constituent quark and antiquark. In the sequence, we analyze the solution above of the Wigner function for $c\overline{c}$ meson. The behavior of the Wigner function for the strong interaction of the heavy quarks are showed in Figs.~(\ref{fig:wigner}-\ref{fig: wigner1}). 
\newline

\begin{table}[!ht]
\centering
\caption{The experimental values for linear confinement $\lambda$, reduced mass $m$, and maximum relative distance $q$ for $c\overline{c}$ meson~\cite{A_1,A_50}}
\begin{tabular}{c c c c} \hline \hline
 (a) & (b) & (c) & (d) \\ [0.5ex] 
 \hline
 $\frac{p^2}{2m}=0$ $MeV$ & $\frac{p^2}{2m} =100$ $MeV$ &  $\frac{p^2}{2m}=200$ $MeV$ & $\frac{p^2}{2m}=400$ $MeV$\\ 
 $\lambda =600$ $MeV^{2}$ (Exp.) & $\lambda =600$ $MeV^{2}$ (Exp.) & $\lambda =600$ $MeV^{2}$ (Exp.) & $\lambda =600$ $MeV^{2}$ (Exp.)\\
 $m = 630$ $MeV$ (Exp.) & $m = 630$ $MeV$ (Exp.) & $m = 630$ $MeV$ (Exp.) & $m = 630$ $MeV$ (Exp.)\\  
 $q=4077$ $MeV^{-1}$ (Exp.)& & \\[1ex] 
 \hline \hline
\end{tabular}
\label{table:1}
\end{table}
\newpage
For the $c\overline{c}$ meson, we are taking into account the Eq. \eqref{G_3}. The $c\overline{c}$ meson is shown in Fig. \ref{fig:wigner} represented by Wigner function for ground state. The experimental parameter, $\lambda$ linear confinement parameter, $m$ reduced mass, $q$ maximum relative distance between quark and antiquark are all in units of $MeV$ from $c\overline{c}$ meson which is shown in Tab. \ref{table:1}. From Fig. \ref{fig:wigner} is observed for ground state of charmonium meson which the Wigner function provides values negative in the phase space. This behavior is related to the character quantum. Then, it was make cut in graph $3D$ as illustrated in Fig. \ref{fig: wigner1}. In Fig. \ref{fig: wigner1} shows that when $p=0$ $MeV$, the Wigner function presents negative values, curve (a). This is because of quantum interference, also visualized in the curves (b), and (c). In particular, $p=0$ $MeV$ corresponds a limit. It is worth noting that when $p=0$ the energy is $E=9.7$ $MeV$. When varying the kinetic energy to $p=100$ $MeV$, the Wigner function is displaced to the left, curve (b). We can keep increasing the kinetic energy for $p = 200 MeV$ , which further evidences the distance for the left of the graph of the Wigner function, curve (c). By varying the kinetic energy for $p=400$ $MeV$, it is noticed that the graph of the Wigner function is moved away visibly to the left side, curve (d). Thus, it is observed that there is a limit to the right for the existence of the Wigner function not to be zero. This limit approaches the experimental value that refers maximum relative distance $q=4077$ $MeV^{-1}$, TAB. \ref{table:1}. Then, a boundary condition is placed at zero. Consequently, the $c\overline{c}$ meson only exists between zero and where Wigner function decay. Note that the analysis in phase space is revealing that for the variation of the kinetic energy there is an upper limit of existence of the charm-anticharm meson, Fig. \ref{fig: wigner1}. Although it is simple to describe the quark case through a linear potential, the confinement is observed, which is not seen with the configuration space solution.

\section{Cornell Potential and Schrödinger equation in Phase Space }\label{galilean}
In this section, we analyze the quark-antiquark interaction in phase space context. For this purpose, we consider the Cornell potential given in Eq.(\ref{cornell1}). In this way, we present a solution for the Schrödinger equation with Cornell potential in phase space and the associated Wigner function. The two-dimensional Hamiltonian for the $c\overline{c}$ meson is given as 
\begin{equation}
    H=\frac{P_{1x}^{2}+P_{1y}^{2}}{2m}+\frac{P_{2x}^{2}+P_{2y}^{2}}{2m}+V(r), \label{R_1}
\end{equation}
where $m$ is reduced mass in units of the $MeV$. In order to solve the Schrödinger equation for this Hamiltonian, the Bohlin transformation is utilized. Then, the Bohlin mapping is defined by \cite{A_31,A_37,A_38}
\begin{equation}
    x+iy=(q^{2}_{1}-q^{2}_{2})+i(2q_{1}q_{2}),
\end{equation}
or
\begin{align}
    x&=q^{2}_{1}-q^{2}_{2}, \label{T}\\
    y&=2q_{1}q_{2}, \label{T1}
\end{align}
by defining
\begin{equation}
    P_{x}+iP_{y}=\frac{p_1+ip_2}{2(q_{1}+iq_{2})}.
\end{equation}
leads to
\begin{align}
    P_{x}&=\frac{p_1q_1+p_2q_2}{2(q^{2}_{1}+q^{2}_{2})}, \label{T2}\\
    P_{y}&=\frac{p_2q_1-p_1q_2}{2(q^{2}_{1}+q^{2}_{2})}. \label{T3}
\end{align}
Substituting Eqs. \eqref{T}, \eqref{T1}, \eqref{T2}, \eqref{T3} in Eq. \eqref{R_1}, leads to the Hamiltonian 
\begin{equation}
    H=\frac{1}{4}\left [ \frac{(p^{2}_{1}+p^{2}_{1})}{2m(q^{2}_{1}+q^{2}_{2})} \right ]+\frac{b}{(q^{2}_{1}+q^{2}_{2})}+a(q^{2}_{1}+q^{2}_{2}).
\end{equation}
Consequently the hypersurface in phase space defined by $H=E$ leads to
\begin{equation}
    \frac{(p^{2}_{1}+p^{2}_{2})}{2m}-4E(q^{2}_{1}+q^{2}_{2})+4a(q^{2}_{1}+q^{2}_{2})^{2}=-4b, \label{R_0}
\end{equation}
where $p_{1}$ and $p_{2}$ are the canonical momenta conjugate. The Eq. \eqref{R_0} can be written as 
\begin{equation}
    \left [ \frac{(p^{2}_{1}+p^{2}_{2})}{2m}-4E(q^{2}_{1}+q^{2}_{2})+4a(q^{2}_{1}+q^{2}_{2})^{2}+4b\right ]\star \psi(q_{1},p_{1},q_{2},p_{2})=0. \label{R_2}
\end{equation}
Observe that the Eq. \eqref{R_2} is obtained from the classical Hamiltonian through of the star product. Therefore, the Bohlin mapping that leads to Eq. \eqref{R_0} is a classical transformation. For this reason, the perturbation theory is used to analyse the Eq. \eqref{R_2}. So, the equation in phase space is given by
\begin{equation}
    \left [ \widehat{H_0}+\widehat{H_1}\right ]\star \psi(q_{1},p_{1},q_{2},p_{2})=-4b\psi(q_{1},p_{1},q_{2},p_{2}),
\end{equation}
where $ \widehat{H_0}=\frac{(p^{2}_{1}\star+p^{2}_{2}\star)}{2}-4E(q^{2}_{1}\star+q^{2}_{2}\star)$ and $\widehat{H_1}=4a(q^{2}_{1}\star+q^{2}_{2}\star )^{2}$. 

The equation for $\widehat{H_0}$ has the form
\begin{equation}
    \widehat{H_0}\psi^{(0)}(q_{1},p_{1},q_{2},p_{2})=b_{n_1,n_2}^{(0)}\psi^{(0)}(q_{1},p_{1},q_{2},p_{2}), \label{Z}
\end{equation}
where $\psi^{(0)}(q_{1},p_{1},q_{2},p_{2})$ corresponds the eigenfunction the unperturbed Hamiltonian.

Defining the operators
\begin{align}
\widehat{a}&=\sqrt{\frac{mW }{2k}}\left ( q_{1}\star+\frac{ip_{1}\star}{mW } \right ), \\
\widehat{a}^{\dagger}&=\sqrt{\frac{mW }{2k}}\left ( q_{1}\star-\frac{ip_{1}\star}{mW } \right ), \\
\widehat{b}&=\sqrt{\frac{mW }{2k}}\left ( q_{2}\star+\frac{ip_{2}\star}{mW } \right ), \\
\widehat{b}^{\dagger}&=\sqrt{\frac{mW }{2k}}\left ( q_{2}\star-\frac{ip_{2}\star}{mW } \right ).
\end{align}
Where $-4E=\frac{mW^2}{2}$, and the star operators $q_{i}\star$, $p_{i}\star$ are given by
\begin{align}
    q_{i}\star&=q_{i} + \frac{i}{2}\frac{\partial }{\partial p_i}, \\
    p_{i}\star&=q_{i} - \frac{i}{2}\frac{\partial }{\partial q_i},
\end{align}
so, the Hamiltonian is
\begin{equation}
     \widehat{H}=\frac{(p^{2}_{1}\star+p^{2}_{2}\star)}{2}-4E(q^{2}_{1}\star+q^{2}_{2}\star)+4a(q^{2}_{1}\star+q^{2}_{2}\star )^{2}.
\end{equation}
The unperturbed Hamiltonian is defined as
\begin{equation}
     \widehat{H_0}=\frac{(p^{2}_{1}\star+p^{2}_{2}\star)}{2}-4E(q^{2}_{1}\star+q^{2}_{2}\star),
\end{equation}
and the perturbed part is
\begin{equation}
     \widehat{H_1}=4a(q^{2}_{1}\star+q^{2}_{2}\star )^{2}.
\end{equation}
The equation that is to be analyzed is given by
\begin{equation}
    \widehat{H}\star \psi(q_{1},p_{1},q_{2},p_{2})=-4b\psi(q_{1},p_{1},q_{2},p_{2}). \label{p}
\end{equation}
The unperturbed equation is
\begin{equation}
    \widehat{H_0}\psi^{(0)}(q_{1},p_{1},q_{2},p_{2})=b_{n_1,n_2}^{(0)}\psi^{(0)}(q_{1},p_{1},q_{2},p_{2}). \label{Z_1}
\end{equation}
The unperturbed part $\widehat{H_0}$ has solutions given by
\begin{equation}
\psi_{n_1,n_2}^{0}(q_{1},p_{1},q_{2},p_{2})=\Phi _{n_1}(q_1,p_1)\Gamma _{n_2}(q_2,p_2) ,   
\end{equation}
where $\Phi _{n_1}(q_1,p_1)$ and $\Gamma _{n_2}(q_2,p_2)$ are solutions. The acting operators in the states vectors is
\begin{align}
    \widehat{a}\Phi _{n_1}=\sqrt{n_1}\Phi _{n_1-1}, \label{w} \\
    \widehat{a}^{\dagger}\Phi _{n_1}=\sqrt{n_1+1}\Phi _{n_1+1}, \label{w1}\\
   \widehat{ b}\Gamma  _{n_2}=\sqrt{n_2}\Gamma  _{n_2-1},\label{w3} \\
   \widehat{ b}^{\dagger}\Gamma  _{n_2}=\sqrt{n_2+1}\Gamma  _{n_2+1}.\label{w4}
\end{align}
Utilizing the relations $\widehat{a}\Phi _{0}=0$, $\widehat{b}\Gamma  _{0}=0$, the ground state solution is
\begin{equation}
    \psi_{0,0}^{(0)}(q_1,p_1,q_2,p_2)=Ne^{-(q_{1}^2+p_{1}^2)}L_{n_1}(q_{2}^{2}+p_{2}^{2})e^{-(q_{2}^2+p_{2}^2)}L_{n_2}(q_{2}^{2}+p_{2}^{2}), \label{h}
\end{equation}
where $L_{n_1}$, $L_{n_2}$ are Laguerre polynomials and $N$ is a normalisation constant. The eigenvalue solutions obtained of Eq. \eqref{Z_1} are 
\begin{equation}
    b_{0,0}^{(1)}=\frac{kW }{4}(n_{1}+n_{2}+1)+\frac{4ak^2}{mW^2}.
\end{equation}
The states excited are obtained from Eq. \eqref{h} using operators given in Eqs. \eqref{w}, \eqref{w1}, \eqref{w3}, \eqref{w4}. So, the solution for the first-order perturbed is given by
\begin{align}
    \psi^{(1)}_{n_1, n_2}(q_1,p_1,q_2,p_2)&=\psi^{(0)}_{n_1,n_2}(q_1,p_1,q_2,p_2) \nonumber\\
    &+\sum_{\substack{m_1\neq n_1,\\m_2 \neq n_2}}\frac{\langle \psi^{(0)}_{m_1,m_2}(q_1,p_1,q_2,p_2)|\widehat{H_1}|\psi^{(0)}_{n_1,n_2}(q_1,p_1,q_2,p_2)\rangle}{b^{(0)}_{n_1,n_2}-b^{(0)}_{m_1,m_2}}  \label{R_5} \\
    &\times \psi^{(0)}_{m_1,m_2}(q_1,p_1,q_2,p_2).  \nonumber
\end{align}
Note that is required to solve the following expression before a solution for Eq. \eqref{R_5}.

\begin{equation}
    I=\langle \psi^{(0)}_{m_1,m_2}(q_1,p_1,q_2,p_2)|\frac{a}{W^2}\left[ (a^{\dagger}+a)^{2}+(b^{\dagger}+b)^{2} \right]^{2} |\psi^{(0)}_{n_1,n_2}(q_1,p_1,q_2,p_2)\rangle. 
\end{equation}
Using the orthogonality relations
\begin{align}
    \langle \Phi _{n}^{*}(q_{1},p_{2})|\Phi _{m}(q_{1},p_{1})\rangle=\delta _{n,m}, \\
     \langle \Gamma  _{n}^{*}(q_{1},p_{2})|\Gamma  _{m}(q_{1},p_{1})\rangle=\delta _{n,m},
\end{align}
the ground state of $c\overline{c}$ meson is
\begin{equation}
    \psi^{(1)}_{0,0}=\psi^{(0)}_{0,0}+\frac{a}{2W^3}\left [ -4\sqrt{2}\psi^{(0)}_{2,0}-\psi^{(0)}_{2,2}-\frac{7}{2}\sqrt{2}\psi^{(0)}_{0,2}-\frac{\sqrt{6}}{2}\psi^{(0)}_{0,4} \right ].
\end{equation}

\begin{figure}[htbp]
\subfigure[Wigner function for ground state $n=0$ of the $c\overline{c}$ meson.\label{m1}]{
\includegraphics[scale=0.32]{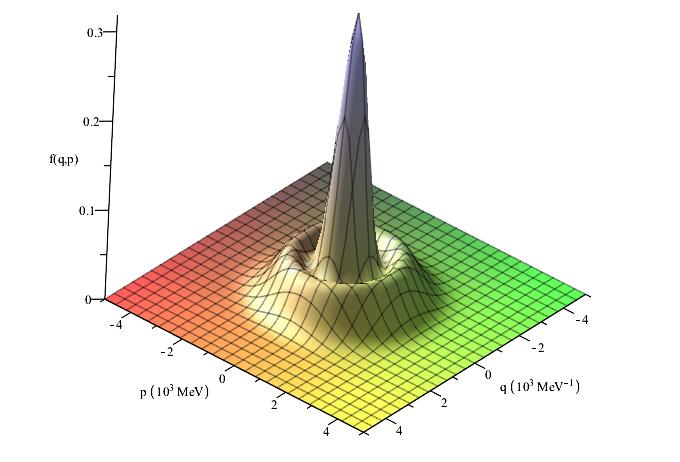}}
\subfigure[Wigner function for first excited state $n=1$ of the $c\overline{c}$ meson.\label{m2}]{
\includegraphics[scale=0.32]{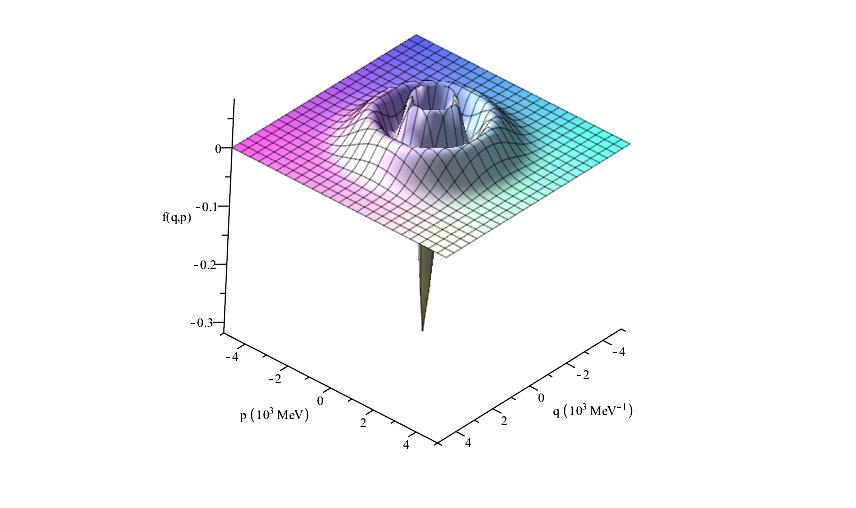}}
\end{figure}

And for excited states of the $c\overline{c}$ meson, the wave functions are
\begin{equation}
    \psi^{(1)}_{1,0}=\psi^{(0)}_{1,0}+\frac{a}{2W^3}\left [ -\frac{\sqrt{30}}{2}\psi^{(0)}_{5,0}+\frac{-1-11\sqrt{6}}{2}\psi^{(0)}_{3,0}-3\psi^{(0)}_{3,2}-5\sqrt{2}\psi^{(0)}_{1,2}-\frac{\sqrt{6}}{2}\psi^{(0)}_{1,4} \right ],
\end{equation}
and
\begin{equation}
    \psi^{(1)}_{0,1}=\psi^{(0)}_{0,1}+\frac{a}{2W^3}\left [ -\frac{\sqrt{30}}{2}\psi^{(0)}_{0,5}+\frac{-1-11\sqrt{6}}{2}\psi^{(0)}_{0,3}-3\psi^{(0)}_{2,3}-5\sqrt{2}\psi^{(0)}_{2,1}-\frac{\sqrt{6}}{2}\psi^{(0)}_{4,1} \right ].
\end{equation}
The Wigner function for the $c\overline{c}$ meson is given by
\begin{equation}
    f_{w}(q_{1},p_{1},q_{2},p_{2})=\psi^{(1)}_{n_1,n_2}(q_{1},p_{1},q_{2},p_{2})\star \psi^{\dagger(1)}_{n_1,n_2}(q_{1},p_{1},q_{2},p_{2}).
\end{equation}
To obtain the corrections of the first order of energy is requiring to solve the matrix
\begin{eqnarray}
W=\begin{pmatrix}
W_{aa} & W_{ab}\\ 
W_{ba} & W_{bb}
\end{pmatrix},
\end{eqnarray}
where $W_{ij}=\langle \psi_{i}^{(0)}|H_{1}|\psi_{j}^{(0)} \rangle$. Hence we get
\begin{eqnarray}
W=\begin{pmatrix}
\frac{25ak^2}{m^2W^2} & \frac{2ak^2}{m^2W^2}\\ 
\frac{17ak^2}{m^2W^2} & \frac{25ak^2}{m^2W^2}
\end{pmatrix}.
\end{eqnarray}
The eigenvalue obtained are $\lambda _{1}=30,83\left ( \frac{ak^2}{m^2W^2} \right )$ and $\lambda _{2}=19,17\left ( \frac{ak^2}{m^2W^2} \right )$. Then, the corrections of the first order of eigenvalue of the Eq. \eqref{p} is
\begin{align}
    b_{1,0}^{(1)}=\frac{-kW}{4}(n_{1}+n_{2}+1)+30,83\left ( \frac{ak^2}{m^2W^2} \right ), \\
    b_{0,1}^{(1)}=\frac{-kW}{4}(n_{1}+n_{2}+1)+19,17\left ( \frac{ak^2}{m^2W^2} \right ).
\end{align}
Fig. \ref{m1} and Fig. \ref{m2} shows plots of the Wigner function associated to the ground state and first excited state for $c\overline{c}$ meson. It should be noted that all plots consider the coordinates $q_{2}$, $p_{2}$ constant in order to show a three-dimensional figure, while $q_{1}=q$, $p_{1}=p$. In Fig. \ref{m2}, one clearly sees the presence of negative values for the Wigner function at the level excited of the $c\overline{c}$ meson. In Fig. \ref{m1}, the Wigner function looks strictly positive to the ground state.

With the Wigner function determined for the $c\overline{c}$ meson, we can calculate the negativity parameter for this system, this parameter is correlated to the nonclassicality of the system~\cite{kenfack2004}. The results of this calculation are shown in table \eqref{table:2}. As can be seen in table \eqref{table:2}, the negativity parameter increases when $n_{1}$, $n_{2}$ grow. The negativity parameter is a relevant tool in the context of Wigner formalism because one is related to non-classicality of physical systems as soon as is also related to chaotic behavior. In this sense, analysing the Wigner function negativity  it is possible study aspects which the usual wavefunction formalism does not allows. Then, the study in phase space is crucial in order to understand more about the chaotic nature.  

\begin{table}[!htbp]
\centering
\caption{Negativity parameter $\eta (\psi )$ for the levels $n_{1}, n_{2}=0,1$.}
\scalebox{1.4}{%
 \begin{tabular}{||c c c c c c c c c ||} 
 \hline
$n_{1},n_{2}$&\qquad&\qquad&\qquad\qquad&\qquad\qquad&\qquad\qquad&\qquad\qquad&\qquad&$\eta(\psi)$\\ [1ex] 
\hline
 0,0 &\qquad&\qquad&\qquad\qquad&\qquad\qquad&\qquad\qquad&\qquad\qquad&\qquad\qquad& 0\\
0,1&\qquad&\qquad&\qquad\qquad&\qquad\qquad&\qquad\qquad&\qquad\qquad&\qquad\qquad& 0 \\
1,0&\qquad&\qquad&\qquad\qquad&\qquad\qquad&\qquad&\qquad\qquad&\qquad\qquad& 0.230\\
 \hline
\end{tabular}} \label{table:2}
\end{table}

\newpage
\section{Concluding Remarks}\label{sec:Conclusion}
In this paper, we have studied heavy quarkonium bound states using the Schrödinger-like equation with an approach the Cornell potential in the framework Symplectic Quantum Mechanics. In particular, this sector is characterized by a linear potential which leads to the quarks confinement plus one Coulomb-type potential. This is called of Cornell potential that feature experimentally this sector of quantum chromodynamics QCD. Initially, the Wigner function was obtained for the $c\overline{c}$ meson system. It was observed that analyzing in the phase space allow for simple model to observe the mesons confinement why it is not seen into configuration space solution. We check out through the analysis in the phase space that due to the variation of kinetic energy, it has a maximum limit of the existence to the mesons systems given by graphics obtained. After that, we have obtained the energy eigenvalues and the respective wave functions using perturbation theory. Such that the states, denoted a quasi-amplitude of probability, is associated with Wigner function by the Weyl product. In this context, using the Wigner functions, was obtained the negativity parameter for the ground state and first excited state of $c\overline{c}$ meson system. Finally, we observe through the analysis in the phase space that due to the change of energy levels, the negativity parameter increasing Tab. \ref{table:2}. This is show that existing the behaviour nonclassical for these system. It was not seen in the literature recent works.

\section{Acknowledgments}

I would like to thank CNPq and CAPES of Brazil by support. 

\end{document}